# GHz Laser-free Time-resolved Transmission Electron Microscopy: a Stroboscopic High-duty-cycle Method

Jiaqi Qiu, Gwanghui Ha,[*] Chunguang Jing, Sergey V. Baryshev[†]
Euclid TechLabs, 365 Remington Blvd., Bolingbrook, IL 60440, USA

Bryan W. Reed
Integrated Dynamic Electron Solutions, 5653 Stoneridge Dr., Suite 117, Pleasanton, CA 94588, USA

Yimei Zhu
Department of Condensed Matter Physics and Materials Science, Brookhaven National Laboratory, Upton, NY 11973, USA

June W. Lau
Materials Science and Engineering Division, National Institute of Standards and Technology, Gaithersburg, MD 20899, USA

**Abstract.**

A device and a method for producing ultrashort electron pulses with GHz repetition rates via pulsing an input direct current (dc) electron beam are provided. The device and the method are based on an electromagnetic-mechanical pulser (EMMP) that consists of a series of transverse deflecting cavities and magnetic quadrupoles. The EMMP modulates and chops the incoming dc electron beam and converts it into pico- and sub-pico-second (100 fs to 10 ps) electron pulse sequences (pulse trains) at >1 GHz repetition rates, as well as controllably manipulates the resulting pulses, and ultimately leads to no electron pulse phase-space degradation compared to the incoming dc beam parameters. The temporal pulse length and repetition rate for the EMMP are both continuously tunable in wide ranges.

Applying the EMMP to a transmission electron microscope (TEM) with *any* dc electron source (e.g. thermionic, Schottky, or field-emission source), a GHz stroboscopic high-duty-cycle TEM can be realized. Unlike in many recent developments in time-resolved TEM that rely on a sample pumping laser paired with a laser launching electrons from a photocathode to probe the sample, there is no laser in the presented experimental set-up. This is expected to be a significant relief for electron microscopists who are not familiar with laser systems. The EMMP and the sample are externally driven by a radiofrequency (RF) source synchronized through a delay line. With no laser pumping the sample, the problem of the pump laser induced residual heating/damaging the sample is eliminated. As many RF-driven processes can be cycled indefinitely, sampling rates of 1-50 GHz become accessible. Such a GHz stroboscopic TEM would open up a new paradigm for *in situ* and *in operando* experiments to study samples externally driven electromagnetically. Complementary to the lower (MHz) repetition rates experiments enabled by laser photocathode TEM, new experiments in the high rep-rate multi-GHz regime will be enabled by the proposed RF design. Because TEM is also a platform for various analytical methods, there are infinite application opportunities in energy and electronics to resolve charge (electronic and ionic) transport, and magnetic, plasmonic and excitonic dynamics in advanced functional materials. In addition, because the beam duty-cycle can be as high as ~10%, detection can be accomplished by any number of commercially available detectors.

---

[*] Also with POSTECH, Korea
[†] s.baryshev@euclidtechlabs.com / sergey.v.baryshev@gmail.com

In this article, we report an optimal design of the EMMP and an analytical generalized matrix approach in the thin lens approximation, along with detailed beam dynamics taking actual realistic dc beam parameters in a TEM operating at 200 keV.

## 1. Introduction

The past 10 years has seen enormous advancement in time-resolved transmission electron microscopes (TEMs), driven largely by advances in pulsed laser systems. Time-resolved ultrafast TEM (UTEM) [1] and dynamic TEM (DTEM) [2] are steadily establishing new measurement capabilities to observe, understand and control solid or soft materials at the most basic level under equilibrium, far-from-equilibrium, extreme, or *in situ* and *in operando* conditions. In a broader context, UTEM and DTEM are electron methods to probe structural dynamics, and even though electrons are a factor of $10^5$ stronger in scattering power with matter compared with photons, ultrafast science instrumentation is currently dominated by expensive synchrotron x-ray and x-ray free electron laser facilities, as well as by the use of rare and highly customized high harmonic generation laser microscopes. Since electron- and photon-class tools are fundamentally different, a comparison can be drawn using a combined resolution metric called space-time resolution (STR, spatial resolution times temporal resolution) – in both classes STR can be $\leq 10^{-20}$ m·s. Nevertheless, there are two other parameters that are important: the sampling rate (repetition rate) and beam duty cycle in electron or photon pulse sequences. At the moment, none of the aforementioned techniques/tools is able to provide GHz-scale sampling rates.

Factors limiting sampling rates at light sources are complex, and even if they are solved, the problems of equipment cost and limited beam time allocations will still limit the practical accessibility of GHz-scale methods. In contrast, UTEM and ultrafast electron diffraction [3] approaches can be developed at the scale of single principal investigators, and in this case the barrier to GHz-scale sampling arises from the inherent limitations of driving processes with a pump laser. UTEM systems typically operate at much less than 0.1 GHz, and sometimes even at ~0.1 MHz, depending on the experiment. While lasers with higher repetition rates are available, experiments must nonetheless operate at lower frequencies because the sample, pumped by the laser, must cool and return to its initial state between pulses. In typical UTEM operating conditions, the sample temperature can rise by ~50K in the absorption of a single pulse [4-6]. In contrast to most laser-driven processes, many processes driven electrically/magnetically or both can be cycled indefinitely at GHz frequencies. A high sampling rate technique (high rep-rate combined with large duty cycle) enables previously intractable experiments because it significantly enhances signal-to-noise ratio (SNR). For example, a UTEM with a 100 MHz laser with 10 fs pulse yields a $10^{-6}$ duty cycle, which poses significant SNR challenges even for modern single-electron-sensitive electron detectors. In the Poisson-noise-limited regime, data throughput is directly proportional to duty cycle, and as a result many potential UTEM experiments are currently impractical purely because the sample will not remain stable for the very long exposure times required.

As an alternative to the laser-photocathode combination method of producing electron pulses, blanking of a direct current (dc) electron beam can produce periodic electron pulse sequences with a flexible temporal structure that can be perfectly synchronized with the clock signal driving a high-frequency nanoscale device (be it a transistor, a plasmonic laser diode, a nano-electromechanical system (NEMS), a spin-transfer torque memory, or any other such device). Such a microscope would reveal the inner workings of such devices in unprecedented ways, by bringing all of the high-resolution imaging, analytical, nanoscale diffraction (including strain measurements), and other capabilities (such as holographic imaging of electric fields or spectroscopic imaging of plasmonic fields combined with tomography, making it truly 4D) of a modern TEM down to the time scale of the device's normal operation. In the method being proposed, the pulsed laser system is eliminated, which significantly reduces system complexity and enhances the ease of use. These are important factors to consider when studying already highly-complicated systems for high-technology applications.

The first prototypes of time-resolved electron microscopes operating in a stroboscopic regime were created back in the 1960s. Those were scanning electron microscopes (SEM) used to image electron current propagation in novel (for that time) semiconductor devices such as MOSFETs [7] and Gunn-effect devices [8]. The achievable frame size (single pulse length) and frame (repetition) rate parameters were driven by progress in semiconductor technology – from MHz in 1968 [7] to GHz in 1978 [8]. Thus, by early 1980s stroboscopic SEMs and TEMs looking into processes with temporal resolution as low as 10 ps were widely prototyped. While ns-pulses at MHz repetition rates were easy to generate using a standard deflecting plate system [7], ps-pulses at GHz repetition rates required developments of new strategies, which included a specialty meander travelling-wave line [9], specialty fast capacitors [10], and an RF pillbox cavity with a deflecting mode operating at 1 GHz applied for practical use in electron microscopy for the first time [8]. The aforementioned examples made use of SEMs with lateral resolution just below 1 micron and very early-generation TEMs with spatial resolution much worse than in current instruments. While imaging with spatial resolution between 1 Å and 1 nm is routine in a modern TEM, recent interest in time-resolved and pump-probe experiments sets much more challenging requirements for electron microscopy: additional temporal resolution between 10 ns and <100 fs is now expected on top of the same spatial resolution range.

None of the existing time-resolved microscopes have simultaneously achieved all three of the following parameters: (1) Continuously tunable pulse duration between ~100 fs and ~100 ps; (2) Continuously tunable rates of repetition from ~0.1 GHz to ~50 GHz; (3) Phase-space fidelity of electron pulses so that processes at nanometer length scales and below can be resolved. In this paper, we describe a design of a compact, versatile, drop-in EMMP for medium electron energies (~100 keV) to enable laser-free GHz stroboscopic TEM. The final design was chosen from a number of candidates based on the results of analytical geometrical optics approximation and detailed beam dynamics calculations. The solution is such that when the EMMP is ON, only electrons matching the strobe frequency are allowed access to the sample and the TEM operates in the GHz stroboscopic mode. When the EMMP is OFF, the TEM operates in its conventional dc beam mode.

## 2. Approaches and methods: basic design of the EMMP

To pulse the dc beam, mechanical or electromagnetic approaches can be used. Both mechanical shaping and/or electromagnetic manipulations of electron beams are common in the field of particle accelerators. They are used to tailor electron beam properties and for electron beam diagnostics, including emittance exchange experiments [11], phase-space imaging [12], energy-chirp compensation [13] and accelerator-based radiation sources [14, 15]. Following previous reports on combined electromagnetic-mechanical GHz pulsers [8, 16], the core component consists of a transverse deflecting cavity (TDC) and a chopping collimating aperture (CCA). A concept of this device is shown in Fig.1. A vacuum TDC, externally driven by an RF source, operated in TM110 mode at $f_0$=10 GHz (corresponding to a diameter of 39 mm) and a CCA are used to form ultrahigh repetition rate pulse sequences (the repetition rate here is 20 GHz because pulses are produced by cutting both sides, leading and trailing, of the sinusoid). At the fixed fundamental TDC frequency of 10 GHz, the pulse length can be changed between 100 fs and 10 ps by varying the CCA diameter and/or RF power in the cavity. The exact range of duty cycle depends on the ratio of the diameters of the TDC (determining $f_0$) and the CCA, and the power fed by the RF source into the TDC. For the TM110 mode in a pillbox, a general relation between all the parameters involved is described as

$$P \propto B^2 = \frac{r \times m_e}{d \times e \times \Delta t}, \quad (1)$$

where $P$ and $B$ are power and magnetic component of the electromagnetic field in the TDC, respectively; $m_e$ and $e$ are the electron mass and charge, respectively; $r$ is the radius of the CCA; $d$ is the free-drifting distance between the TDC and the CCA; and $\Delta t$ is the electron pulse length. This leads to duty cycles of up to 20%. The high duty cycle means only modest tweaks to the beam current and/or the detector are needed to obtain an image. One important note: the RF pillbox cavity technology is downwards compatible to sampling rate (or strobe rate) in the MHz by replacing vacuum in the TDC with a high permittivity dielectric. The general relation linking the TDC diameter ($D$), the fundamental TDC frequency ($f_0$) and the permittivity ($\varepsilon$) is $D \sim \dfrac{1}{f_0 \times \sqrt{\varepsilon}}$. With a high permittivity ferroelectric, the TDC can be continuously tunable too in a different frequency range.

The incoming longitudinal electron dc beam traveling along the $z$-axis (Fig.1), blue straight segment originating before $z=0$ cm) picks up a transverse sinusoidal momentum by the electromagnetic (EM) field in the TDC. Since the EM field oscillates with the radial frequency $f_0$, the modulation force depends on the time at which electrons arrive in the TDC. The amplitude of the sinusoid grows as the modulated beam propagates. The CCA is placed on axis to chop the beam and converts it into a pulse sequence. As mentioned in earlier publications on the stroboscopic pulsers/choppers, different scenarios of spatial deterioration of the electron beam were observed [8, 16]. Indeed, using the concept in Fig.1 the beam would expand in the transverse (beam diameter growth) and longitudinal (temporal coherence deterioration) directions after drifting a certain distance. While this was not an issue for imaging on the micron scale , much more rigorous schemes must be considered in order to ensure beam spatial and energy coherence necessary for TEM spectroscopy, diffraction and imaging. Thus, additional optical elements are required downstream of the CCA to mitigate the effects produced by the first TDC.

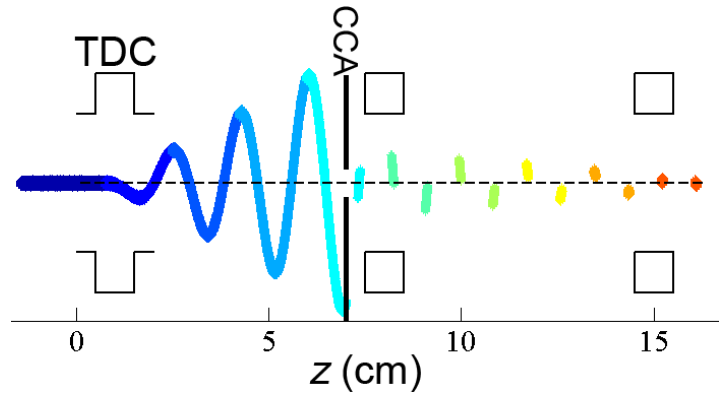

Fig.1. General concept of the compact electromagnetic-mechanical pulser (EMMP) for pulsing dc electron beams and manipulating electron pulses in the phase space. (The $z$-scale is approximate, and color coding is arbitrary.)

## 3. Generalized matrix calculations

Since the final goal is integration with a TEM, the EMMP should be as compact as possible. This not only places an upper limit on the number of optical elements, but also makes it desirable to minimize the number of necessary adjustments to tune the EMMP. To come up with the best miniature design for removing the post-TDC distortions of the resulting electron pulses, we use a generalized matrix approach. In this approach, optical elements are described by matrices. Matrix components depend on the actual geometry of the EMMP, and the strength of various effects on electron dynamics in the phase space can be predicted and evaluated. The matrix methodology relies on three basic assumptions: (1) electron optics elements are approximated as thin lenses; (2) it considers a single particle/electron; (3) only linear matrix transformations are considered. All three assumptions are intertwined. When combined, they establish the basis for the geometrical optics framework in which the problem is solved. This idealized framework provides a good first-order model for rapid progress in the design, to be followed up with full ray-trace calculations including space charge effects to determine the effects of aberrations and undesired couplings on the electron phase space.

An initial and a final state of an electron at input and at the output of the EMMP are linked in the momentum-coordinate phase space via a beam transport matrix as follows

$$\begin{pmatrix} x_f \\ x'_f \\ y_f \\ y'_f \\ z_f \\ \frac{\Delta p_f}{p_0} \end{pmatrix} = \begin{pmatrix} R_{11} & R_{12} & R_{13} & R_{14} & R_{15} & R_{16} \\ R_{21} & R_{22} & R_{23} & R_{24} & R_{25} & R_{26} \\ R_{31} & R_{32} & R_{33} & R_{34} & R_{35} & R_{36} \\ R_{41} & R_{42} & R_{43} & R_{44} & R_{45} & R_{46} \\ R_{51} & R_{52} & R_{53} & R_{54} & R_{55} & R_{56} \\ R_{61} & R_{62} & R_{63} & R_{64} & R_{65} & R_{66} \end{pmatrix} \begin{pmatrix} x_i \\ x'_i \\ y_i \\ y'_i \\ z_i \\ \frac{\Delta p_i}{p_0} \end{pmatrix}, \quad (2)$$

where $x$ is the relative horizontal beam position, $x'$ is the horizontal divergence, $y$ is the relative vertical beam position, $y'$ is the vertical divergence, $z$ is the relative longitudinal position (principal optic axis of the TEM) or time, and $\frac{\Delta p}{p_0}$ is the relative longitudinal momentum.

In Eq.2, the matrix $R(6\times6)$ is called the transport matrix. It is a result of multiplication of all matrices describing every single component of an EMMP design; this includes the drifting matrix describing empty gaps/pipes between hardware components. The perfect case is when the matrix $R$ has only diagonal elements: it means an electron beam transformation took place, yet cross-correlations, described by off-diagonal elements resulting in pulse size change in transverse and longitudinal directions and energy spread, are absent. A number of combinations and designs were analyzed, and the main conclusion was that at least 3 active elements must be present to minimize off-diagonal elements. While the first element is always a TDC, second and third elements are shown as blank squares in Fig.1, and should be determined from the matrix analysis. It was found that best results can be anticipated when combinations of TDCs and/or magnetic quadrupoles (MQs) are used, provided such combinations are also most compact and simple from the operational point of view. Two most attractive designs are illustrated in Fig.2.

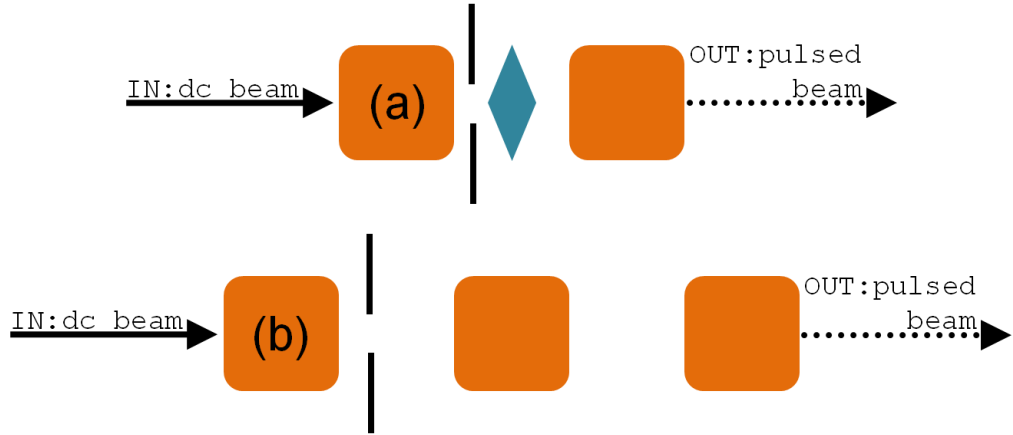

Fig.2. Two main EMMP designs for further considerations based on the matrix analysis (not to scale). The orange square is the TDC, and the cyan diamond is the MQ. Design (a) is abbreviated as TDC+MQ+TDC and design (b) is abbreviated as 3TDC.

Thus, the resulting matrix $R(6\times6)$ is the product of 5-fold multiplication of 3 key matrices related to the EMMP design in Fig.2. Free drift beam pipe of length $d$ (empty space between either optical component in the EMMP) [measured in meters] is described by the following matrix

$$\begin{pmatrix} 1 & d & 0 & 0 & 0 & 0 \\ 0 & 1 & 0 & 0 & 0 & 0 \\ 0 & 0 & 1 & d & 0 & 0 \\ 0 & 0 & 0 & 1 & 0 & 0 \\ 0 & 0 & 0 & 0 & 1 & \dfrac{d}{\gamma^2} \\ 0 & 0 & 0 & 0 & 0 & 1 \end{pmatrix}, \quad (3)$$

where $\gamma$ is the Lorentz factor. Its value depends on the electron energy. The magnetic quadrupole with a focal length $f$ [measured in meters] is described as

$$\begin{pmatrix} 1 & 0 & 0 & 0 & 0 & 0 \\ \dfrac{1}{f} & 1 & 0 & 0 & 0 & 0 \\ 0 & 0 & 1 & 0 & 0 & 0 \\ 0 & 0 & -\dfrac{1}{f} & 1 & 0 & 0 \\ 0 & 0 & 0 & 0 & 1 & 0 \\ 0 & 0 & 0 & 0 & 0 & 1 \end{pmatrix}. \quad (4)$$

The TDC has a matrix

$$\begin{pmatrix} 1 & 0 & 0 & 0 & k & 0 \\ 0 & 1 & 0 & 0 & 0 & 0 \\ 0 & 0 & 1 & 0 & 0 & 0 \\ 0 & 0 & 0 & 1 & 0 & 0 \\ 0 & 0 & 0 & 0 & 1 & 0 \\ k & 0 & 0 & 0 & 0 & 1 \end{pmatrix}, \quad (5)$$

where $k$ is the transverse momentum acquired by an electron in the TDC, and it is measured in reciprocal meters. In what follows, $k$ will be referred to as *kick*. Resulting transport matrices for the designs, sketched in Fig.2, were optimized in order to zero as many off-diagonal elements as possible. For this calculation, the following dc beam input parameters were considered: (1) beam energy ($E_0$) 200 keV; (2) energy spread ($\Delta E$) 0.5 eV; (3) emittance 1.5 nm×rad which was a product of a beam diameter of 10 μm and a divergence angle of 0.15 mrad; these values are also typical for various accelerator applications. The TDC+MQ+TDC design (Fig. 2a) has a transport matrix

$$\begin{pmatrix} -\dfrac{d_2}{d_1} & 0 & 0 & 0 & 0 & 0 \\ \dfrac{(\gamma^2 - d_1^2 k_1^2)(d_1 + d_2)}{d_1 d_2 \gamma^2} & -\dfrac{d_1}{d_2} & 0 & 0 & 0 & \dfrac{d_1 k_1 (d_1 + d_2)}{d_2 \gamma^2} \\ 0 & 0 & 2+\dfrac{d_2}{d_1} & 2(d_1 + d_2) & 0 & 0 \\ 0 & 0 & \dfrac{d_1 + d_2}{d_1 d_2} & 2+\dfrac{d_1}{d_2} & 0 & 0 \\ \dfrac{k_1(d_1 + d_2)}{\gamma^2} & 0 & 0 & 0 & 1 & \dfrac{d_1 + d_2}{\gamma^2} \\ 0 & 0 & 0 & 0 & 0 & 1 \end{pmatrix}, \quad (6)$$

where $d_1$ and $d_2$ are the drift distances between the first TDC and the MQ, and between the MQ and the second TDC respectively; $k_1$ is the kick strength of the first deflecting cavity. The focal length of the MQ is $f = \dfrac{d_1 d_2}{d_1 + d_2}$, and the kick strength of the second deflecting cavity is $k_2 = \dfrac{d_1}{d_2} k_1$. For the 3TDC (Fig.2b) design, the $R$-matrix is

$$\begin{pmatrix} 1-\dfrac{d_1(d_1+d_2)k_1^2}{\gamma^2} & d_1+d_2 & 0 & 0 & 0 & -\dfrac{d_1(d_1+d_2)k_1}{\gamma^2} \\ -\dfrac{d_1(d_1+d_2)k_1^2}{d_2\gamma^2} & 1+\dfrac{d_1^2(d_1+d_2)k_1^2}{d_2(d_1(d_1+d_2)k_1^2-\gamma^2)} & 0 & 0 & 0 & -\dfrac{d_1^2(d_1+d_2)^2k_1^3}{d_2(d_1^2k_1^2+d_1d_2k_1^2-\gamma^2)} \\ 0 & 0 & 1 & d_1+d_2 & 0 & 0 \\ 0 & 0 & 0 & 1 & 0 & 0 \\ 0 & -\dfrac{d_1(d_1+d_2)k_1}{\gamma^2} & 0 & 0 & 1-\dfrac{d_1(d_1+d_2)k_1^2}{\gamma^2} & \dfrac{d_1+d_2}{\gamma^2} \\ 0 & -\dfrac{d_1^2(d_1+d_2)^2k_1^3}{d_2(d_1^2k_1^2+d_1d_2k_1^2-\gamma^2)} & 0 & 0 & -\dfrac{d_1(d_1+d_2)k_1^2}{d_2} & 1+\dfrac{d_1^2(d_1+d_2)k_1^2}{d_2(d_1(d_1+d_2)k_1^2-\gamma^2)} \end{pmatrix}, \quad (7)$$

where $k_2 = -\dfrac{d_1+d_2}{d_2}k_1$ and $k_3 = \dfrac{d_1}{d_2}k_1$ are found as optimal for the overall system design, i.e. maximum off-diagonal elements are zeros.

While the matrix methodology is widely used in high energy physics (electron beam energy range ~10 MeV), it is challenging to apply it to the medium energy electron beams (~100 keV) found in TEMs due primarily to the Lorentz factor. Many off-diagonal elements, in the transport matrices (Eq. 6 and 7) are proportional to $\dfrac{1}{\gamma^2}$. So for a 10 MeV beam where $\gamma \approx 20$, the off-diagonal elements become negligibly small, this is certainly not the case for a 100 keV beam where $\gamma \approx 1$. From the matrices, it is seen that in transverse directions both designs may lead to satisfactory results such that sufficient spatial coherence in the beam is conserved upon EMMP installation. The main problem here is the matrix term $R_{65}$ which is responsible for energy spread growth. In this idealized geometrical optics framework, $R_{65}$ is zero for the TDC+MQ+TDC design, but is finite for the 3TDC design. That is, in the 3TDC case, the additional energy spread at $E_0=200$ keV is higher than 1 eV on top of the default/intrinsic energy spread of 0.5 eV.

## 4. Beam dynamics simulations

Matrix calculations provided a good starting point for the actual EMMP design. In the 3TDC case, it is seen that the energy spread is proportional to the kick value to the second power, $k_1^2$. Since the resulting electron pulse length is controlled by both $k_1$ and the CCA radius, to achieve a short pulse we must either have a large kick or a small aperture. Thus, to reduce energy spread for a given pulse size, we should choose a CCA radius as small as reasonably possible, thus allowing us to reduce the kick. At the same time, if the beam radius is larger than the CCA radius the EMMP will not be able deliver effective chopping at the CCA no matter how strong the kick is. Thus, the radius of the incoming dc beam produces a limitation for the EMMP in terms of the relation between the CCA radius and the kick strength, and consequently a limitation for minimal extrinsic/additional energy spread achievable upon pulsing. Beam dynamics calculations were carried out to compare transverse and longitudinal phase spaces between the input dc beam and the output pulsed beam. Using the geometry and beam characteristics described by Reed *et al.* [17] and keeping in mind the previously described limitations (dc beam radius versus CCA radius), we conducted our simulation in the *soft focusing regime* meaning that an input dc beam radius was 3 μm and an angle of convergence was 0.1 mrad, resulting in beam emittance 0.3 nm×rad. Changing the beam emittance between the default TEM mode and the soft focusing regime can be achieved by inserting a custom condenser lens (C0) similar to one described in Ref.[17], thus allowing a large fraction of the current emitted from the gun to be collected, collimated, and aligned into the pulser with a small convergence angle. Total dc current at C0 of ~100 nA is feasible in this alignment. The soft focusing is illustrated in Fig.3. Dotted red lines represent a way the dc beam should propagate and converge through the EMMP when it is OFF. Using C0 to create a soft crossover at the CCA plane allows optimal modulation and chopping upon turning the EMMP ON. Normal TEM operation with a low-emittance dc beam is obtained by turning off both C0 and the pulser components.

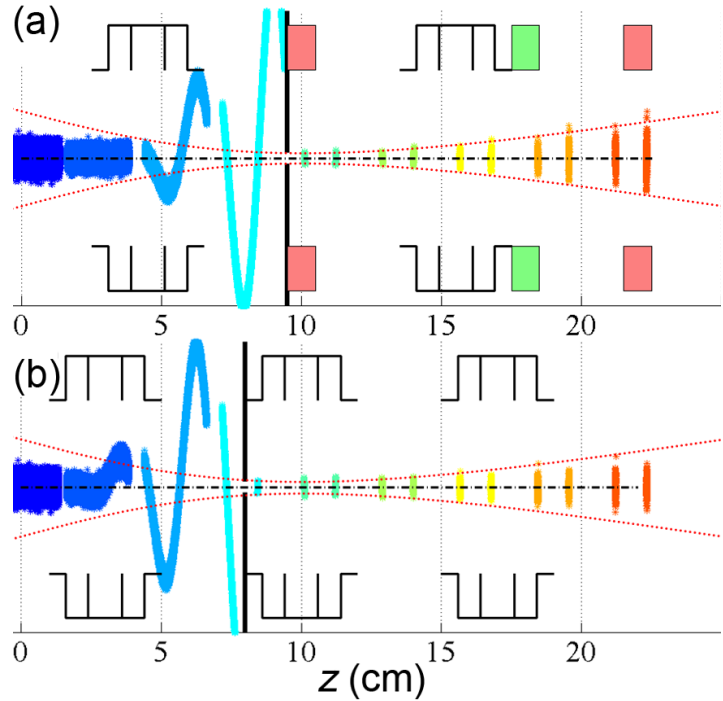

Fig.3. Two key designs identified by the matrix calculations and tested through beam dynamics. (a) TDC+MQ+TDC design; (b) 3TDC design. Red Dotted lines represent beam waist in the middle of the second active EMMP element, i.e. MQ (a) and TDC (b) – not to scale. The central black dash-dotted line is the symmetry (optic) axis.

In terms of the device placement, a combined section containing C0 and EMMP should be inserted after the ultrahigh vacuum gun chamber but before the condenser lens section. This means operation and maintenance under the moderate column vacuum conditions which (1) completely circumvents the need to disturb the electron source and (2) preserves the factory packaging and design of the condenser system. After reviewing microscope ray-diagrams, we found this was the best location for preserving all the available imaging and diffraction modalities. First, we emphasize that there is no change in dc beam propagation before the condenser lens system. Since beam phase-space manipulations to achieve different imaging modes in a TEM column take place at or below the condensers, by inserting the device between the gun and the condenser lenses, the problem is reduced to preserving the transverse and longitudinal phase spaces as the dc beam converts into a pulse train as it passes through the EMMP. Based on the soft focusing parameters, an input file representing the incoming dc beam was built using the code PARMELA [18, 19]. Input dc parameters and geometry used in the calculations are summarized in Table 1 and Fig.3, respectively. Particle tracking through the EMMP in the longitudinal and transverse phase spaces was performed by a custom-developed MATLAB [19] code. Resulting parameters are compiled in Table 1. While both designs deliver nearly perfect conservation of the transverse phase space ($\varepsilon_x$, $\varepsilon_y$), energy spread becomes a major parameter for design considerations. In TDC+MQ+TDC design, an energy spread as low as 0.25 eV was achieved. Given a typical intrinsic energy spread in a TEM of 0.5 eV, the total value is 0.56 eV for a 10 ps pulse. As it was first suggested by matrix/thin lens approximations in Section 3, the TDC+MQ+TDC configuration (Figs.2a, 3a) is chosen for further hardware development based on this analysis. Such an energy resolution will allow pulsed electron energy-loss spectroscopy in TEM.

**Table 1.** Key design quantities (kick and CCA radius), and values characterizing the transverse and longitudinal phase spaces of the pulsed beam at the output of the EMMP with the two basic designs (3TDC and TDC+MQ+TDC) considered.

| EMMP design | pulse length (ps) | *kick* (keV) | aperture radius (µm) | $\varepsilon_x$ (nm×rad) | $\varepsilon_y$ (nm×rad) | energy spread (eV) |
|---|---|---|---|---|---|---|
| INPUT | --- | --- | --- | 0.3 | 0.3 | --- |
| TDC+MQ+TDC | 1 | 1.25 | 5 | 0.24 | 0.39 | 0.81 |
| | 1 | 1.88 | 7.5 | 0.31 | 0.40 | 1.38 |
| | 1 | 2.50 | 10 | 0.38 | 0.43 | 2.23 |
| | 1 | 5.00 | 20 | 0.41 | 0.43 | 9.7 |
| | 10 | 0.25 | 10 | 0.36 | 0.40 | 0.25 |
| | 10 | 0.50 | 20 | 0.38 | 0.39 | 1.01 |
| 3TDC | 1 | 2.5 | 10 | 0.34 | 0.42 | 28.4 |
| | 1 | 5 | 20 | 0.38 | 0.39 | 118 |
| | 10 | 0.5 | 20 | 0.38 | 0.43 | 12.5 |

In the TDC+MQ+TDC design, the MQ can be assembled with electro- or permanent magnets. Making use of a permanent magnet MQ would simplify the overall EMMP implementation. Taking it one step further, let us consider dc beam behavior when the EMMP is OFF. Fig.4 illustrates the root-mean-square (rms) transverse beam size, i.e. beam envelop in two transverse directions, with the EMMP ON (solid red and blue lines) and OFF (dotted red and blue lines). The state OFF means that only the TDCs are OFF, and the MQs are permanent magnets, i.e. ON all the time. It is seen that envelop and divergence after the aperture, especially at the outlet of the EMMP, are nearly identical. This result suggests that to return back to default TEM imaging mode with dc beam, in the TDC+MQ+TDC environment with the permanent magnets the RF source can be simply turned OFF such that TDCs become additional free drift space. If compact electromagnetic quadrupoles are in use, they can be either turned OFF such that the whole EMMP becomes a drift pipe and the dc beam is adjusted via C0, or the electromagnets are additionally used to steer and focus the dc beam, or they stay ON at the nominal values such that the situation in Fig.4 takes place.

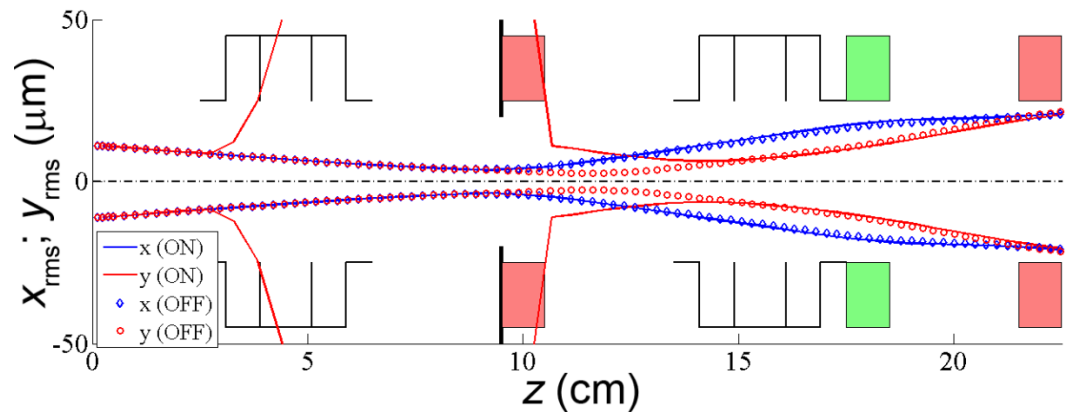

Fig.4. Beam envelop with the EMMP ON (solid lines) and OFF (dotted lines) – not to scale. Apparent discontinuity in the red solid curve is the result of RF modulation of the dc beam. The central black dash-dotted line is the symmetry (optic) axis.

## 5. Conclusions

To summarize, a novel time-resolved GHz stroboscopic concept for transmission electron microscopy is proposed. Such concept is laser-free and is aimed at resolving sub-nanosecond processes in advanced magnetic, electronic, ionic and photonic materials under actual operation conditions. A sample is given a GHz excitation with a radiofrequency source which is phase-locked through a delay line to an RF cavity pulser, which converts a standard dc beam in a TEM into pulse trains being driven by the same radiofrequency source. In the family of time-resolved electron probe methods, the laser-free/electromagnetically-driven GHz stroboscopic concept fulfills a different temporal landscape that is complementary to the existing commercial solutions. We used an analytical matrix algorithm and beam dynamics simulations to identify and rationalize a design-of-choice for a compact pulser for a 200 keV electron beam compatible with TEM electron columns. Table 2 summarizes key performance parameters anticipated in the proposed GHz stroboscopic TEM.

**Table 2.** Summary of parameters considered in the paper.

| | | |
|---|---|---|
| initial beam energy | 200 keV | |
| intrinsic energy spread | 0.5 eV | |
| dc beam current at gun exit | ~100 nA | |
| operation mode | stroboscopic | |
| laser | not required | |
| pulse length | 1 ps | 10 ps |
| repetition rate at specimen | 20 GHz | 20 GHz |
| duty cycle | 2% | 20% |
| number of electrons per cycle | ~1 | ~10 |
| rms emittance | ≤0.4 nm×rad | ≤0.4 nm×rad |
| extrinsic rms energy spread | 0.81 eV | 0.25 eV |
| total energy spread | 0.95 eV | 0.56 eV |
| relative energy spread | $4.75 \times 10^{-6}$ | $2.80 \times 10^{-6}$ |
| STR | $<10^{-21}$ m·s | $<10^{-20}$ m·s |

## Acknowledgments

Euclid TechLabs work was supported by DOE SBIR program grant No. DE-SC0013121. Y.Z. was supported by the U.S. Department of Energy, Office of Science, Office of Basic Energy Sciences, under Contract No. DE-SC0012704.